\documentclass[showkeys,aps,twocolumn,showpacs,preprintnumbers,amsmath,amssymb,prl,letterpaper,floatfix,nofootinbib,superscriptaddress,]{revtex4-1}

\usepackage{amsmath,amssymb}
\usepackage[usenames,dvipsnames]{color}
\usepackage{graphicx}
\usepackage{subfigure}
\usepackage{amssymb}
\usepackage{soul}
\newcommand\bef{\begin{figure}}
\newcommand\eef[1]{\label{fg:#1}\end{figure}}
\newcommand\beq{\begin{equation}}
\newcommand\eeq[1]{\label{#1}\end{equation}}
\newcommand\bet{\begin{table}}
\newcommand\eet[1]{\label{tb:#1}\end{table}}
\newcommand\fgn[1]{Figure \ref{fg:#1}}

\newcommand\tbn[1]{Table \ref{tb:#1}}

\begin{document}
\title{Lattice QCD study of doubly-charmed strange baryons}
\author{Nilmani\ \surname{Mathur}}
\email{nilmani@theory.tifr.res.in}
\affiliation{Department of Theoretical Physics, Tata Institute of Fundamental
         Research,\\ Homi Bhabha Road, Mumbai 400005, India.}

\author{M.\ \surname{Padmanath}}
\email{Padmanath.M@physik.uni-regensburg.de}
\affiliation{Institut  fur  Theoretische  Physik,  Universitat  Regensburg,
Universitatsstrase  31,  93053  Regensburg,  Germany.}


\begin{abstract}
  We present the energy spectra of the low lying doubly-charmed baryons using lattice quantum chromodynamics.  We precisely predict the ground state mass of the charmed-strange $\Omega_{cc} (1/2^{+})$ baryon to be 3712(11)(12) MeV which could well be the next doubly-charmed baryon to be discovered at the LHCb experiment at CERN. We also predict masses of other doubly-charmed strange baryons with quantum numbers $3/2^{+}, 1/2^{-}$, and $3/2^{-}$. 
\end{abstract}

\maketitle

The recent discovery of a doubly-charmed baryon, $\Xi^{++}_{cc} (ccu)$ with a mass of $3621.40\pm0.78$ MeV and lifetime $0.256^{+0.024}_{-0.022} \pm 0.014$ {\it{ps}}
by the LHCb Collaboration \cite{Aaij:2017ueg,Aaij:2018wzf} marks an important milestone in heavy hadron physics. Consistency in the prediction on the
mass of this state from several lattice calculations~\cite{Lewis:2001iz,Mathur:2002ce,Liu:2009jc,Briceno:2012wt,Basak:2012py,Basak:2013oya,Namekawa:2013vu,Brown:2014ena,Bali:2015lka,Padmanath:2015jea,Chen:2017kxr,Alexandrou:2017xwd,Mondal:2017nhw}, including ours~\cite{Basak:2012py,Basak:2013oya,Padmanath:2015jea,Mondal:2017nhw}, and potential model studies~\cite{Karliner:2014gca} demonstrates 
the depth in our understanding about the theory of quantum chromodynamics (QCD).  Success of theoretical studies in predicting this baryon has boosted the scientific interest in studying the prospects of discovering more doubly 
 heavy hadrons and understanding their properties \cite{Karliner:2018hos,Karliner:2017elp,Karliner:2017qjm,Eichten:2017ffp,Mathur:2018epb,Junnarkar:2017sey,Aaij:2018zrb}. The next obvious doubly-charmed baryon to be searched for is its spin-3/2 partner. Indeed a relatively close $3/2^{+}$ excitation with hyperfine splitting about 80-100 MeV is predicted by various theoretical studies.
Being so closely spaced, its radiative decay to the $1/2^+$ ground state is expected to dominate which makes it difficult for LHCb~\cite{Aaij:2014jba} to observe this particle in near future. However, $\Omega_{cc}(ccs)(1/2^+)$, the strange analogue of $\Xi_{cc}(1/2^{+})$, could well be observed soon at LHCb through its weak decay.
As discussed recently in Ref.~\cite{Karliner:2018hos}, LHCb may be in good position to detect this excitation in decay modes such as $\Xi_c^{0}K^{-}\pi^{+}\pi^{+}$ and $\Omega_{c}\pi^{+}$.
Therefore a timely precise prediction of the ground state mass of $\Omega_{cc}(1/2^+)$ is highly expected.
In this work we perform such a calculation using lattice QCD,
 and make precise predictions of the mass of this baryon as well as masses of its excitations with spin parity $1/2^{-}, 3/2^{+}$, and $3/2^{-}$.

The charm quark being heavy, lattice calculations of charmed hadrons, particularly 
with multiple charm quarks, are plagued by the ultraviolet cut-off effects (lattice spacing). Thanks to recent developments in algorithms and accessibility of petaflops computing, gauge 
ensembles at multiple fine lattice spacings and adequate lattice volumes are available providing opportunities
to perform detailed investigations of charmed hadrons on the lattice~\cite{Briceno:2012wt,Basak:2012py,Basak:2013oya,Namekawa:2013vu,Brown:2014ena,Bali:2015lka,Padmanath:2015jea,Chen:2017kxr,Alexandrou:2017xwd,Mondal:2017nhw,Padmanath:2013zfa,Mathur:2018epb}. In this work taking advantage of such a set of three gauge ensembles generated with lattice spacings of about $0.12, 0.09$ and $0.06$ fermi, we perform a detailed calculation to extract the ground state masses of $\Omega_{cc}$ baryons with $J^{P} = {1\over 2}^{\pm}$ and ${3\over 2}^{\pm}$.
A combination
of various novel tools such as wall source (to obtain better signal-to-noise ratio for the ground state), overlap fermions
(with no ${\mathcal{O}}(ma)$ errors) and judicious utilization of mass differences as well as dimensionless ratios for controlled continuum extrapolations through multiple lattice spacings make our calculation scrupulous in details compared to any other previous such study. This enable us to predict $\Omega_{cc}$ baryons most precisely to date that can be tested at LHCb and/or other future charm facilities. Below we elaborate the numerical details.

\section{\bf{Numerical details}}
\noindent{\bf{A. Lattice ensembles: }}
We perform this calculation on three dynamical 2+1+1 flavours ($u/d,s,c$) lattice ensembles generated by the MILC collaboration~\cite{Bazavov:2012xda}.
These  ensembles, with lattice sizes  $24^3 \times 64$, $32^3 \times 96$ and $48^3 \times 144$, at gauge couplings $10/g^2 = 6.00, 6.30$ and $6.72$, respectively, are generated with the Highly Improved Staggered Quarks (HISQ) action and with the one-loop Symanzik gauge action. The lattice spacings as measured using the $r_1$
parameter for the set of ensembles used here are 0.1207(11), 0.0888(8) and 0.0582(5) fm, respectively
\cite{Bazavov:2012xda}.

\noindent{\bf{B. Quark actions: }}
For valence quark propagators, we employ the overlap fermion action \cite{Neuberger:1997fp,Neuberger:1998wv}, which has exact chiral symmetry at finite lattice spacings~\cite{Neuberger:1997fp, Neuberger:1998wv, Luscher:1998pqa}
and is automatically $\mathcal{O}(ma)$ improved for all flavors.
We utilize wall sources on Coulomb gauge fixed lattices
 to generate light to charm quark propagators.

\noindent {\textbf{C. Quark mass tuning: }
The effects of discretization is the dominating systematic in the lattice 
study of heavy hadrons and crucially depends on the tuning of heavy quark masses. We follow the Fermilab prescription of heavy quarks~\cite{ElKhadra:1996mp} and tune the charm quark mass
by equating the spin-averaged kinetic mass of the $1S$ charmonia 
($\bar{M}_{kin}(1S) = {3\over 4} M_{kin}(J/\psi) + {1\over 4} M_{kin}(\eta_c)$) to its experimental value~\cite{Patrignani:2016xqp}.  
 The tuned bare charm quark masses are found to be 0.290, 0.427 and 0.528
 on fine to coarse lattices respectively, all of which satisfy $m_ca << 1$.
 Following Ref. \cite{Chakraborty:2014aca}, the strange quark mass is tuned by equating the lattice estimate of $\bar{s} s$ pseudoscalar
to 688.5 MeV \cite{Basak:2012py,Basak:2013oya}.

\noindent {\textbf{D. Hadron interpolators: }
We use the conventional baryon interpolators given by 
$P^{\pm}[(q_1^T C\Gamma q_2) q_3]; \Gamma=\gamma_5$ or $\gamma_i, i \equiv x,y,z $ (discussed in detail in Refs. \cite{Mathur:2002ce, Lewis:2001iz, Brown:2014ena}).
Here first two quarks within parenthesis could be $(cc)$ or $(cs)$ diquarks. The first one follows from a non-relativistic 
Heavy Quark Effective Theory (HQET) picture while the later is relativistic~\cite{Bali:2015lka}.  The 
$[(q_1^T C\gamma_{i} q_2) q_3]$ type operator has both spin 1/2 and spin 3/2 components and at zero momentum its correlation function is given by~\cite{Benmerrouche:1989uc} 
\beq
C_{ij}(t) = (\delta_{ij}-{1\over 3}\gamma_{ij})C_{3/2}(t) + {1\over 3}\gamma_{ij}C_{1/2}(t).
\eeq{Pij}
We then use respective projection operators to obtain the spin 3/2 and 1/2 parts ($C_{3/2}(t)$ and $C_{1/2}(t)$).

In this work results for spin-1/2 states are obtained from the relativistic $P^{\pm}[(q_1^T C\gamma_5 q_2) q_3]$ type operators. On the other hand, HQET-based interpolators are employed to investigate the effects of heavy 
quark symmetry in doubly charmed baryons. In the heavy
quark limit, the two heavy quarks effectively act as an almost
point-like color-antitriplet heavy diquark source ($\bar{\mathbb{Q}}$)
\cite{Manohar:1992nd}, which then in combination with the light quark
can form a color-neutral hadron similar to a heavy-light meson
($\bar{\mathbb{Q}}s$).  We would like to
mention here that usefulness of HQET-based operators for these baryons really
depends on how heavy is the heavy quark. While for the bottom quark these
might be suitable, the extracted masses for the charmed baryons may be
subjected to HQET corrections.
We will thus utilize only the relativistic interpolators to predict the charm hadron masses,  while the mass estimates from HQET-based
interpolators can provide an estimation of possible relativistic corrections to the HQET-based picture.

\section{Results}
To reduce the systematics associated with
cut-off effects, instead of calculating the hadron masses we extract the
mass differences. This method was found to be very effective previously \cite{Shanahan:1999mv,Mathur:2002ce,Dowdall:2012ab,Brown:2014ena}.
Since the charm quark 
mass is tuned with the spin average $1S$ charmonia mass, $\overline{1S}_{cc}$,
we calculate the mass difference on the lattice as 
\beq
\Delta M_{B,cc} = [M^{L}_{B,cc} - \overline{1S}_{cc}]a^{-1}.
\eeq{submass}
On each lattice we calculate this subtracted mass and then perform the 
continuum extrapolation to get its continuum value $\Delta M^{c}_{B,cc}$. Finally the physical result is obtained by
adding the physical values of spin average mass to $\Delta M^{c}_{B,cc}$ as 
\begin{equation}
M_{B,cc} = \Delta M^{c}_{B,cc} + (\overline{1S}_{cc})_{phys}.
\end{equation}
We also use following dimensionless ratio of the calculated hadron mass to the $1S$ spin average mass,
\begin{equation}
	R_{B,cc} =  {{M^{L}_{B,cc}}\over{\overline{1S}_{cc}}},
\end{equation}
which is then extrapolated to the continuum limit ($R^{c}_{B,cc}$) and the doubly-charmed mass is obtained from
\begin{equation}
M_{B,cc} = R^{c}_{B,cc} \times (\overline{1S}_{cc})_{phys}.
\end{equation}
These procedures of utilizing dimensionless ratios as well as mass differences for the continuum extrapolations  substantially 
reduce the systematic errors arising from cut-off effects and heavy quark mass tuning.  We use both equations (2) and (4) and found consistent results and add the difference in systematics. Below we discuss results for $\Omega_{cc}$  baryons.

To show the robustness of the ground state mass extraction, in \fgn{fig_eff_mass}, we show the commonly used
effective mass plot.
The top figure
represents the effective mass of $\Omega_{cc}(1/2^{+})$ baryon (in
arbitrary units) corresponding to the relativistic operator. A long plateau
covering ten slices with stable fit is observed (pink band). The bottom
figure corresponds to the effective hyperfine splitting obtained from
the ratio of two-point correlators of $3/2^{+}$ and $1/2^{+}$ baryons.
\bef[tbh]
\centering
\includegraphics*[scale=0.40]{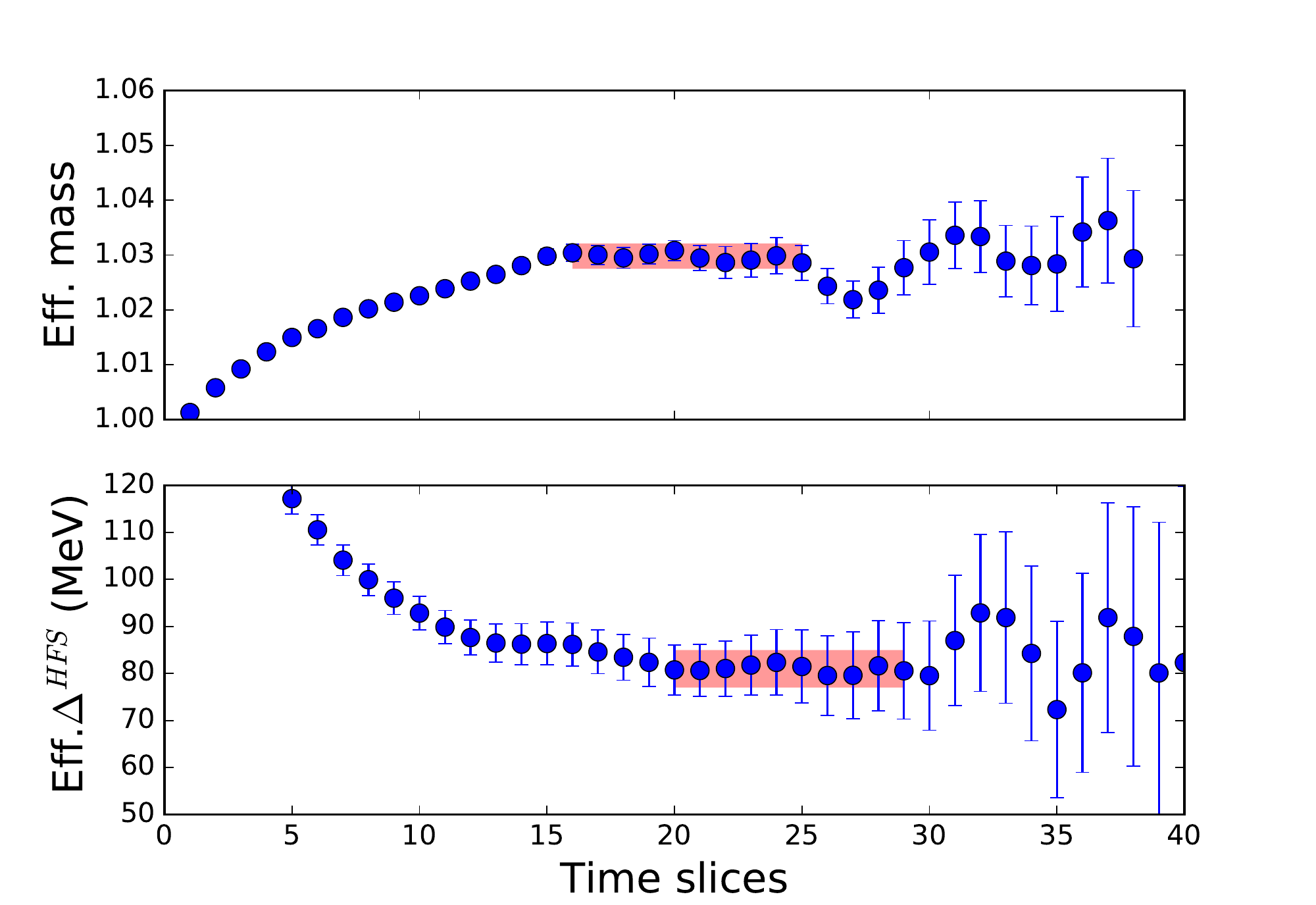}
\vspace*{-0.09in}
\caption{(Top) Effective mass plot of the ground state of $\Omega_{cc}(1/2^{+})$ baryon (in arbitrary units) corresponding to relativistic operator, and (Bottom) effective hyperfine splitting (in units of MeV) on fine lattice.}
\eef{fig_eff_mass}

%
 In \fgn{fig_ccs_split}, we show the results for $\Omega_{cc} (1/2^{+})$ baryons at three
lattice spacings and at the continuum limit (in each plot upper one is for the HQET interpolator and the lower is for the relativistic one).
Top figure corresponds to
the energy splittings (Eq. (2)) while the bottom
one is for the ratio (Eq. (4)). For the continuum extrapolation we use following fit forms : i) $Q(a) = A + a^2B$, ii) $Q_l(a) = A + a^2C\,log(a)$, and iii)  $Cn(a) = A + a^2B+ a^2C\,log(a)$ (with good chiral symmetries (and locality) one would expect that only even powers of $a$ appear, multiplied by coefficients which are polynomial in $log(a)$). With only three data points we perform constrained fits \cite{Lepage:2001ym,Chen:2004gp,McNeile:2012qf} with $Cn(a)$ where the prior values of $B$ and $C$ are constrained with ratio $C_{prior}/B_{prior}$ in between 0.001 to 1 (while varying $B_{prior}$ in a wide range). With the given precision of our dataset, we conclude that it is not possible to quantitatively discern the leading {\it log} term since the deviation between fit results from $a^2$ and leading {\it log} term is negligible in the range of lattice spacings under study. While the quadratic fit form determines the central value of the final result, any difference from it with other fit forms are included as systematic error. In \fgn{fig_ccs_split}, the continuum extrapolated results with the quadratic form are shown by red stars (same symbol and color coding will be used throughout).
Inserting that into Eq. (3), we predict the ground state mass of $\Omega_{cc} (1/2^{+})$ to be 3712(11)(12), obtained from the the relativistic interpolators.
Using ratios (Eq. (4)) we also extract these masses and find results are consistent with the above values.

Note that the HQET interpolator estimates a mass of 3735(11)(12) MeV which is 23 MeV higher than that we obtain from the relativistic interpolator. We believe this difference is due to the non-heaviness of the charm quark (see supplemental materials).
Difference in results indicate strong relativistic corrections to the HQET picture of the $\Omega_{cc}$ baryon, and hence use of such operators may not provide the correct ground state masses for doubly charmed baryons.
It is worthwhile to mention that in our previous investigation \cite{Padmanath:2015jea}, with temporal lattice spacing $a_t \sim 0.035$ fm, we included a large basis of interpolators 
with all different flavor structures allowed for such baryons and followed a detailed 
variational approach.
With optimized operators we found that the ground state mass of the $\Omega_{cc} 
(1/2^{+})$ baryon is 3705(7) MeV,
which is consistent with the mass predictions using the relativistic interpolator.
\bef[tbh]
\centering
\includegraphics*[scale=0.38]{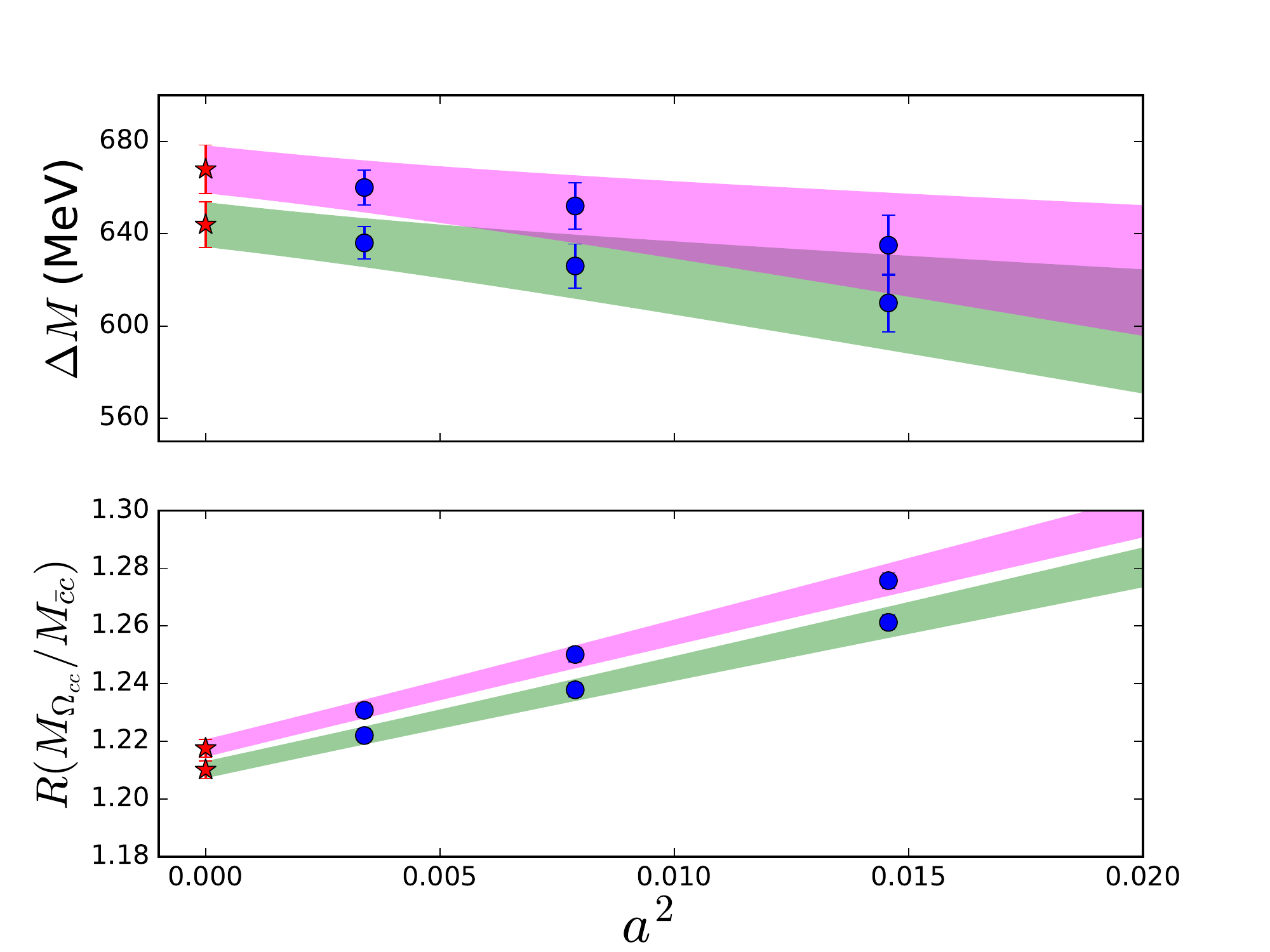}
\vspace*{-0.09in}
\caption{Ground state mass of $\Omega_{cc}(1/2^{+})$ baryon at three lattice spacings are shown in terms of (top) energy splittings from the spin-average mass  (Eq. (2)) and (bottom) the ratio with the spin-average mass (Eq. (4)).  Continuum extrapolated values are also shown by the star symbol. Two cases are for HQET (upper) and relativistic (lower) interpolators.}
\eef{fig_ccs_split}
\bef[tbh]
\vspace*{-0.2in}
\centering
\includegraphics*[scale=0.42]{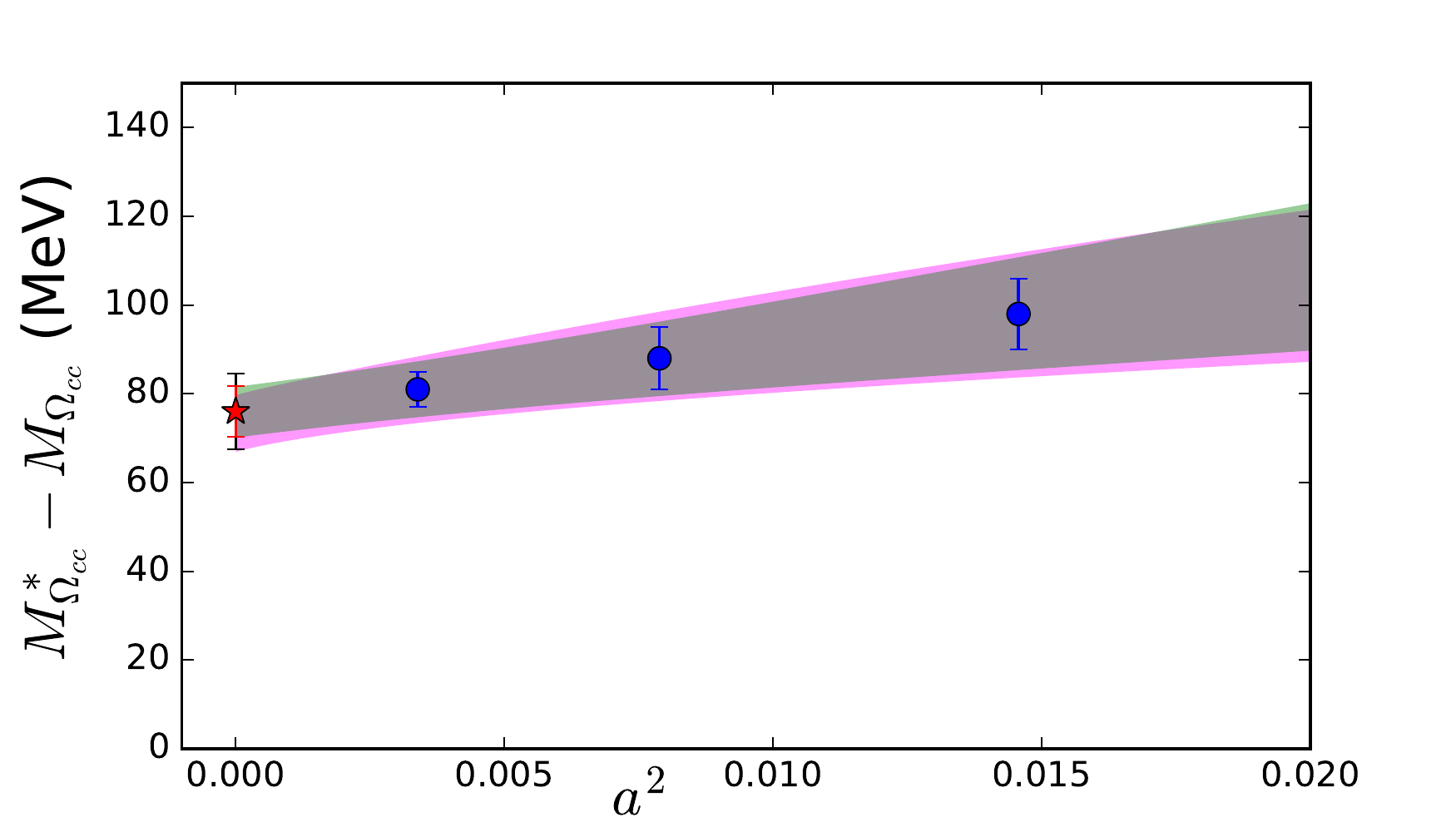}
\caption{Hyperfine splitting between $3/2^{+}$ and $1/2^{+}$ $\Omega_{cc}$ baryons are shown at three lattice spacings and at the continuum limit. Bands represent one sigma-errors in fits with quadratic and logarithmic forms in lattice spacing.}
\eef{fig_hfs_lat}

\bef[tbh]
\centering
\includegraphics*[height=1.5in,width=3.0in]{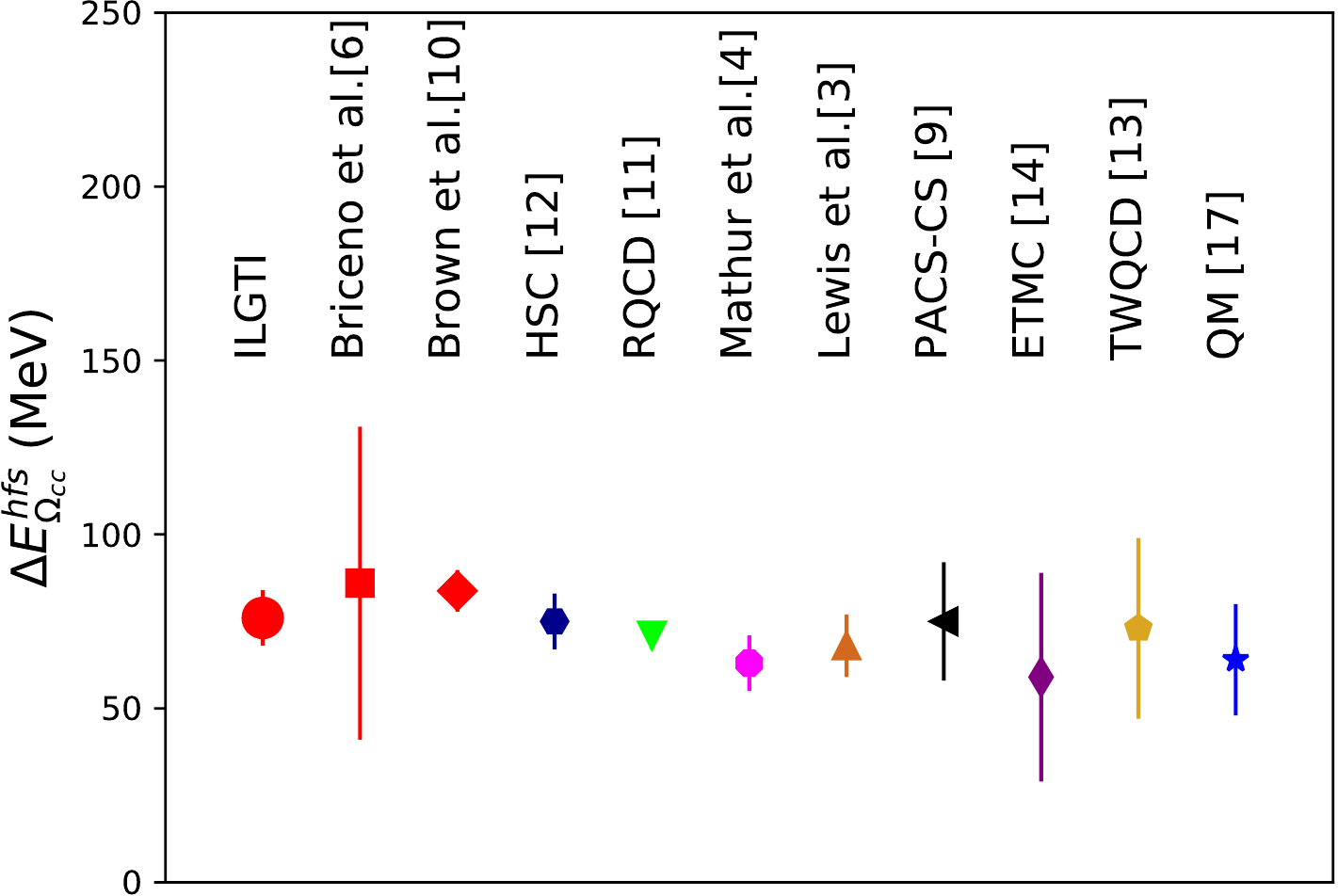}
\vspace*{-0.09in}
\caption{Comparison of hyperfine splitting between the ground states of $3/2^{+}$ and $1/2^{+}$  baryons obtained from various theoretical calculations. Continuum extrapolated results are shown by symbols with red color while symbols with all other colors are obtained only at one lattice spacing.}
\eef{fig_hfs}

In \fgn{fig_hfs_lat} we show the  hyperfine splitting between $3/2^{+}$ and $1/2^{+}$ $\Omega_{cc}$ baryons at three lattice spacings and at the continuum limit (statistical error: red; systematic and statistical errors in quadrature: black lines). 
 Our final result on this hyperfine splitting is 76(6)(6) MeV. In \fgn{fig_hfs}, we summarize the existing lattice results and recent quark model result for this splitting~\cite{Lewis:2001iz,Mathur:2002ce,Liu:2009jc,Briceno:2012wt,Basak:2012py,Basak:2013oya,Namekawa:2013vu,Brown:2014ena,Bali:2015lka,Padmanath:2015jea,Chen:2017kxr,Alexandrou:2017xwd,Mondal:2017nhw,Karliner:2018hos}.

We also calculate the ground state masses of the negative parity baryons. In \fgn{fig_split_n}, we show energy differences (Eq. (2)) of these baryons from the $1S$ spin-average mass.
 Using Eq. (3) we then obtain masses of these baryons as $4071(25)(18)$ and $4112(26)(20)$ for $\Omega_{cc}(1/2^{-})$ and $\Omega_{cc}(3/2^{-})$, respectively, which are consistent with our previous calculation~\cite{Padmanath:2015jea} as well as results from RQCD collaboration~\cite{Bali:2015lka} but smaller than those of Ref.~\cite{Chen:2017kxr}.
\bef[tbh]
\centering
\includegraphics*[scale=0.45]{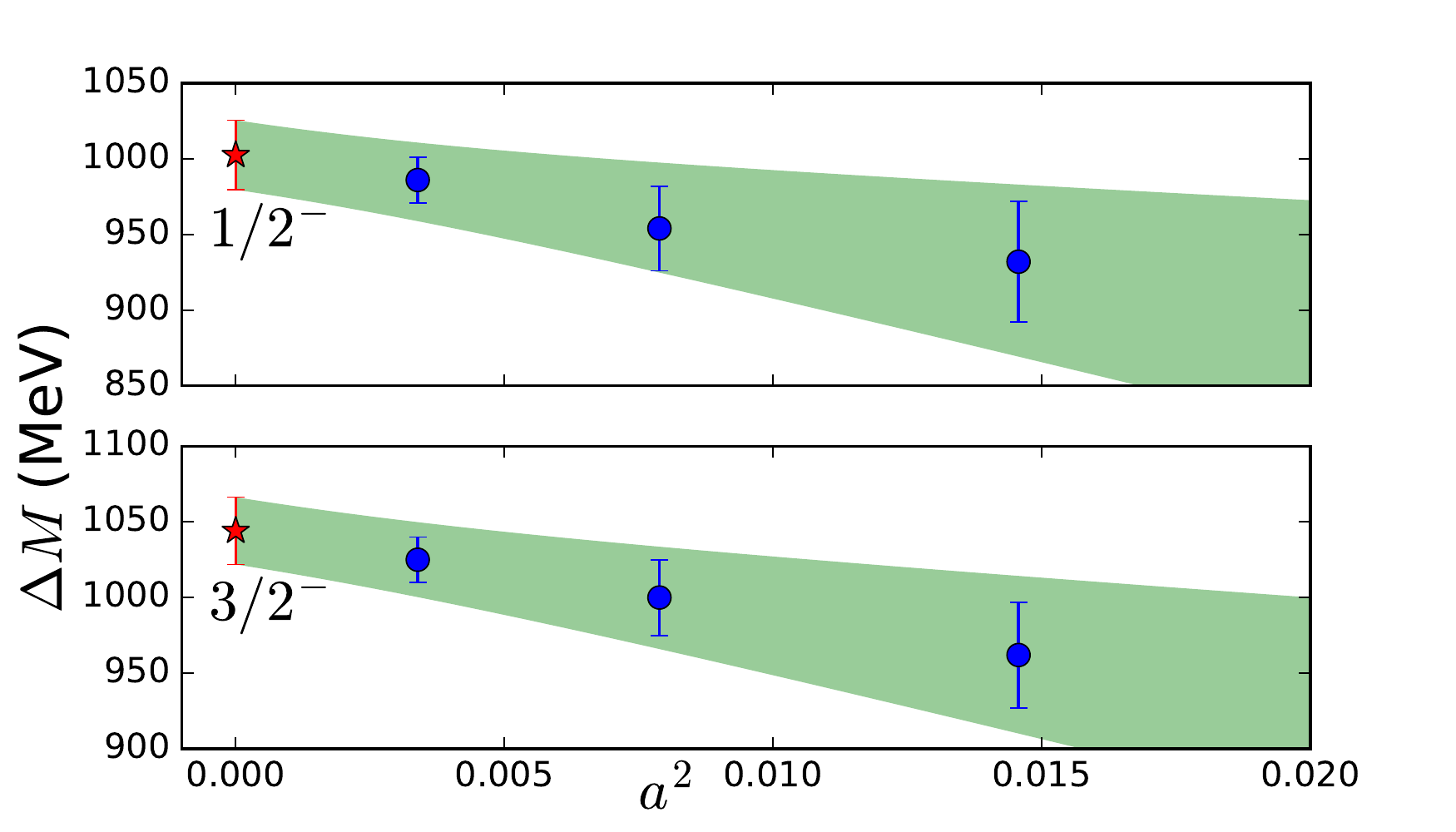}
\vspace*{-0.09in}
\caption{Ground state mass of (top) $\Omega_{cc}(1/2^{-})$ and (bottom) $\Omega_{cc}(3/2^{-})$ at three lattice spacings are plotted in terms of the energy splittings from the spin-average mass (Eq. (2)).}
\eef{fig_split_n}
The relevant strong decay scattering channels that can influence the $1/2^-$ and 
$3/2^-$ masses are $\Xi_{cc}K$ and $\Xi_{cc}^*K$ respectively. However, any quantitative comments on such hadronic  interactions are beyond the scope of this work.
%
\bet[h]
\centering
\begin{tabular}{c|c }
$\Omega_{cc}$ & Lattice Prediction (MeV)\\
\hline
$1/2^{+}$ & 3712(11)(12)  \\
$3/2^{+}$ & 3788(13)(12)  \\
$1/2^{-}$ & 4071(25)(18)  \\
$3/2^{-}$ & 4112(26)(20)  \\
\hline
\end{tabular}
\caption{Low lying $\Omega_{cc}$ baryons as predicted in this work.}
\eet{summary_table}

\bef[tbh]
\centering
\includegraphics*[height=2.4in,width=3.4in]{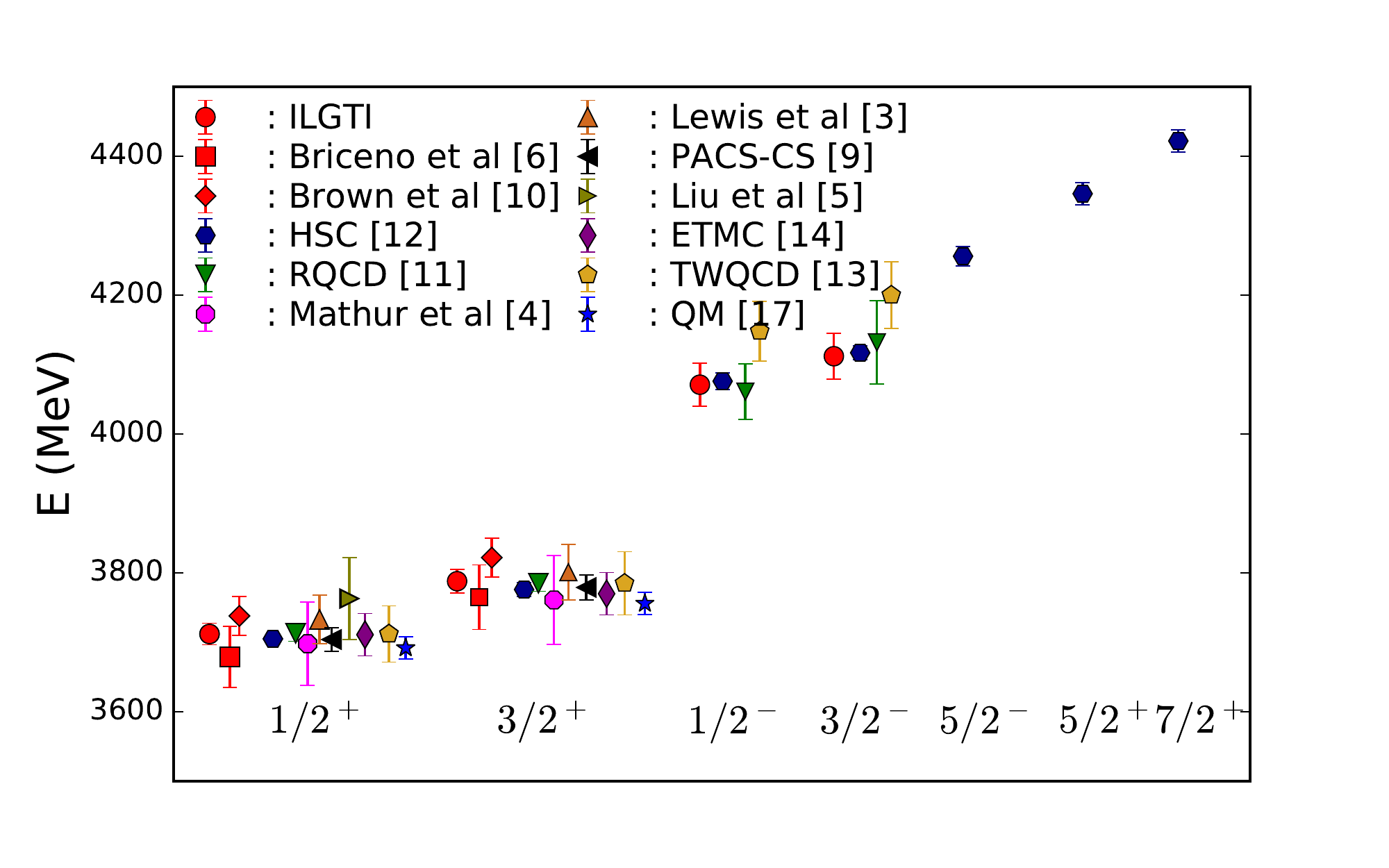}
\vspace*{-0.09in}
\caption{Energy spectra of the low lying $\Omega_{cc}(ccs)$ baryons obtained from different lattice calculations and a recent quark model calculation. Our results are represented as ILGTI (this calculation) and HSC (previous calculation). Continuum extrapolated results are shown by symbols with red color while symbols with all other colors are obtained only at one lattice spacing.}
\eef{fig_summary}

In \tbn{summary_table} we summarize our results, and in \fgn{fig_summary} we show these results (red circles) for all the low-lying doubly-charmed baryons along with other lattice results~\cite{Lewis:2001iz,Mathur:2002ce,Liu:2009jc,Briceno:2012wt,Basak:2012py,Basak:2013oya,Namekawa:2013vu,Brown:2014ena,Bali:2015lka,Padmanath:2015jea,Chen:2017kxr,Alexandrou:2017xwd,Mondal:2017nhw} and a recent quark model calculation~\cite{Karliner:2018hos}. We would like to comment that our results are obtained after controlled continuum extrapolation of the results from three lattice spacings.
 Among the other lattice results only Ref.~\cite{Briceno:2012wt} utilized three lattice spacings but its errors are too big for any precise predictions. The lattice bare charm quark masses ($m_ca$) of Ref.~\cite{Briceno:2012wt} are also much larger compared to those of this calculation, particularly at coarse lattice, and so could well be affected by discretization errors. Results of  Ref.~\cite{Brown:2014ena} are obtained from two lattice spacings and all other results are obtained from only one lattice spacing. 
Below we give error budget for $\Omega_{cc}(1/2^{+})$.
\bet[h]
\centering
\begin{tabular}{l|c }
$Source$ & Error (MeV)\\
\hline
Statistical & 10  \\
 Discretization & 8  \\
  Scale setting& 4   \\
  $m_c$ tuning & 3  \\
  $m_s$ tuning & 4 \\
  Fit window & 1 \\
  Finite size & 3 \\
  Electromagnetism & 3\\
  \hline
  Total & 10 (stat) \& $<$ 12 (syst)\\
  \hline
\end{tabular}
\caption{Error budget in the calculation of $\Omega_{cc}$ baryon.}
\eet{error_table}
 
\noindent{\it{Statistical}}:  The use of wall sources helps to obtain long and stable fit ranges in the correlation functions, 
as demonstrated in \fgn{fig_eff_mass}. We find a statistical uncertainty of 10 MeV for $\Omega_{cc}(1/2^{+})$.\\
{\it Fitting window error}: With long and stable plateau we find uncertainty due to different fitting windows for the 
$\Omega_{cc}(1/2^{+})$ baryon to be about 1 MeV. The largest (4 MeV) fitting window error is found to be for $\Omega_{cc}(3/2^{-})$ baryon. \\
{\it Discretization}:
The use of overlap action ensures no $\mathcal{O}(ma)$ and $\mathcal{O}(ma)^3$ errors. The tuned bare charm quark masses are found to be small ($am<<1$), which assure higher order errors are smaller, particularly at the finest lattice. In addition to that the mass
splittings as well as dimensionless ratios for continuum extrapolations,
ensure reduced discretization errors beyond $\mathcal{O}(ma)$.
 Furthermore, within the acceptable $\chi^{2}$/dof, the
extrapolations are performed using both the quadratic and the logarithmic fit forms, as well as with a constrained fit with both forms together (relevant fitted results are added in supplemental materials). Difference in central values from different
extrapolations are included in discretization errors and altogether we find $<$ 8 MeV uncertainty from discretization.
\\
{\it Scale setting error}: An alternate determination of the lattice spacing was performed \cite{Basak:2013oya} by measuring the $\Omega_{sss}$
baryon mass and were found to be consistent with the determinations using $r_1$ parameter \cite{Bazavov:2012xda}. Measurement of scale with Wilson flow~\cite{Bazavov:2015yea} was also found to be consistent with the scale used here.
The scale setting uncertainty in the mass difference (Eq. (2)) for $\Omega_{cc}(1/2^+)$ is found to be $\sim$ 4 MeV. \\
{\it Charm quark mass tuning error}: The charm quark mass is tuned following the Fermilab prescription \cite{ElKhadra:1996mp}.
Furthermore, the mass splittings being smaller than the masses themselves, the effects due to the mistuning of the charm quark mass are expected to be very small  \cite{Mathur:2002ce,Dowdall:2012ab,Brown:2014ena}. An uncertainty in quark mass tuning is estimated based on interpolating results from multiple 
charm quark mass around the tuned mass. For $\Omega_{cc}(1/2^{+})$, we find this to be $\sim$ 3 MeV. \\
{\it Strange quark mass tuning}: Again, as in the charm quark mass tuning we use multiple strange quark masses and Eqs. (2) and (4) are utilized to see the effect in splittings and ratios. We find a maximum uncertainty of $\sim$ 4 MeV in this mass tuning \\
{\it Finite size effects}: Studies of the same observables on ensembles with similar lattice size indicated finite 
size effects to be within an MeV \cite{Brown:2014ena}. We include an uncertainty of 3 MeV from finite volume effects.\\
{\it Other sources}: For these baryons no chiral extrapolation is involved. Errors due to mixed action effects are found to be small within this lattice set up~\cite{Basak:2014kma}
and are expected to vanish in the continuum limit. The unphysical sea quark mass effects are expected to be within 
a percent for these observables with no effective valence light quark content~\cite{McNeile:2012qf, Dowdall:2012ab, Chakraborty:2014aca}. Errors from electromagnetism are expected  to be within 3 MeV~\cite{Borsanyi:2014jba}. These errors are summarized in \tbn{error_table} and adding all in quadrature we find an overall uncertainty less than 12 MeV.

\noindent{\bf{Conclusions:}} In this Letter, using various state-of-the-art lattice techniques, we present a precise prediction of the ground state mass of $\Omega_{cc}(1/2^{+})$ baryon using lattice QCD with very good control over systematics. We predict the mass of this particle to be 3712(11)(12) MeV. We also predict masses of other $\Omega_{cc}$ baryons with spin-parity quantum numbers $3/2^{+}, 1/2^{-}$ and $3/2^{-}$ to be 3788(13)(12), 4071(25)(18) and 4112(26)(20) MeV, respectively. The hyperfine splitting between the ground state masses of  $3/2^{+}$ and $1/2^{+}$ baryons is found to be 76(6)(6) MeV. Using lattice ensembles at three different lattice spacings, finest one being 0.0582 fermi, we perform a 
controlled continuum extrapolation to determine physical spectra of these baryons. 
A combination of various novel tools like the use of overlap fermions, wall source and prudent utilization of mass differences as well as dimensionless ratios for continuum extrapolations enables us to predict these states very precisely than any previous lattice calculation. Our 
final results for the ground state masses of all $\Omega_{cc}$ baryons are tabulated in Table II and 
also showed in ~\fgn{fig_summary}. We also find that the HQET interpolators are not suitable for these doubly charmed baryons as the charm quark is not so-heavy.


To date only one doubly-charmed baryon, the spin  $\Xi_{cc}(1/2^{+})$, has been discovered. However, given the lattice QCD as well as potential model predictions, the discovery of the spin $\Xi_{cc}^{*}(3/2^{+})$ could be delayed at the LHCb experiment due to its near proximity to  $\Xi_{cc}$. On the other hand the ground state of the $\Omega_{cc}(1/2^{+})$ baryon can possibly be discovered by identifying similar decay channels as that of $\Xi_{cc}$. Our precise prediction in this work could aid such a search for this subatomic particle.

\section{Acknowledgements} We thank our colleagues within the ILGTI collaboration.
We are thankful to the MILC collaboration 
and in particular to S. Gottlieb for providing us with the HISQ lattices. We are thankful to an unknown referee for pointing us on the absence of cubic terms in the continuum extrapolations with chiral fermions. 
Computations are carried out on the Cray-XC30 of ILGTI, TIFR, 
and on the Gaggle/Pride clusters of the Department of Theoretical Physics,
TIFR. N. M. would like to thank Stefan Sint for his valuable comments on continuum extrapolation. N. M. would also like to thank P. Junnarkar for discussions and Ajay Salve, Kapil Gadhiali and P. M. Kulkarni for computational supports. M. P. acknowledges support from EU under grant no. MSCA-IF-EF-ST-744659 (XQCDBaryons) and 
the Deutsche Forschungsgemeinschaft under Grant No.SFB/TRR 55.

\bibliography{Omegacc_arXiv}


\section{Supplementary Materials}
\subsection{Continuum extrapolation of lattice data for doubly charmed strange baryons using different fit forms in lattice spacing ($a$)}
\vspace*{-0.1in}
\bet[h]
\centering
\begin{tabular}{cc||cc}
$a^2$ & $\Delta M (a^2)$ (MeV)&Fit form & $\Delta M (a^2 = 0)$ (MeV)\\
\hline
0.00338724 & 660(7.5)  &  $Q$   & 667.9(10)\\
0.00788544 & 652(10)   &  $Q_l$ & 671.6(12)\\
  0.01456849         &   635(13)        & $Q^{e}_l$ & 670(11)\\
&           & $Q^p_l$ & 668.4(11)\\
\hline
0.00338724 & 636(7)    &  $Q$   & 643.9(9.7)\\
0.00788544 & 626(9.5)  &  $Q_l$ & 647.9(11)\\
0.01456849& 610(12.5)          & $Q^{e}_l$ & 646.5(11)\\
&           & $Q^p_l$ & 644.6 (11)\\

\hline
\end{tabular}
\caption{Mass splittings and fitted results related to Figure 2. Fit forms are $Q: A + a^2 B$ and $Q_{l}: A + a^2 C \,log(a)$, $Q^{e}_{l}: A + a^2 D (1+log(a))$ and $Q^p_{l}: A + a^2 E + a^2 F\,log(a)$. The last fit is a constrained fit with priors taking the ratio $C_{prior}/B_{prior}$ from 0.001 to 0.9 (while varying $B_{prior}$ in a wide range).}
\eet{Fig2}

\vspace*{-0.15in}

\bet[h]
\centering
\begin{tabular}{cc||cc}
$a^2$ & $\Delta M (a^2)$ (MeV)&Fit form & $\Delta M (a^2 = 0) (MeV)$\\
\hline
0.00338724  &  81(4)   & $Q$  & 75.8(5.7)\\
0.00788544  & 88(7) & $Q_l$ &73(6.3)\\
0.01456849&   98(8)        & $Q^{e}_l$ & 74(6.4)\\
&           & $Q^p_l$ & 75(6.4)\\

\hline
\end{tabular}
\caption{Mass splittings and fitted results related to Figure 3.}
\eet{Fig3}

\vspace*{-0.15in}

\bet[h]
\centering
\begin{tabular}{cc||cc}
$a^2$ & $\Delta M (a^2)$ (MeV)&Fit form & $\Delta M (a^2 = 0) (MeV)$\\
\hline
0.00338724 & 986(15)  &  $Q$ & 1002(23)\\
0.00788544 & 954(28) &  $Q_l$ &1011(27)\\
0.01456849& 932(40)          & $Q^{e}_l$ & 1008(26)\\
&           & $Q^p_l$ & 1005(25)\\
\hline
0.00338724 & 1025(15)  &  $Q$ & 1044(22)\\
0.00788544 & 1000(25) &  $Q_l$ &1053(26)\\
0.01456849& 962(35)         & $Q^{e}_l$ & 1050(25)\\
&           & $Q^p_l$ & 1047(25)\\
\hline
\end{tabular}
\caption{Mass splittings and fitted results related to Fig 5.}
\eet{Fig5}


\newpage

\bef[tbh]
\centering
\includegraphics*[scale=0.35]{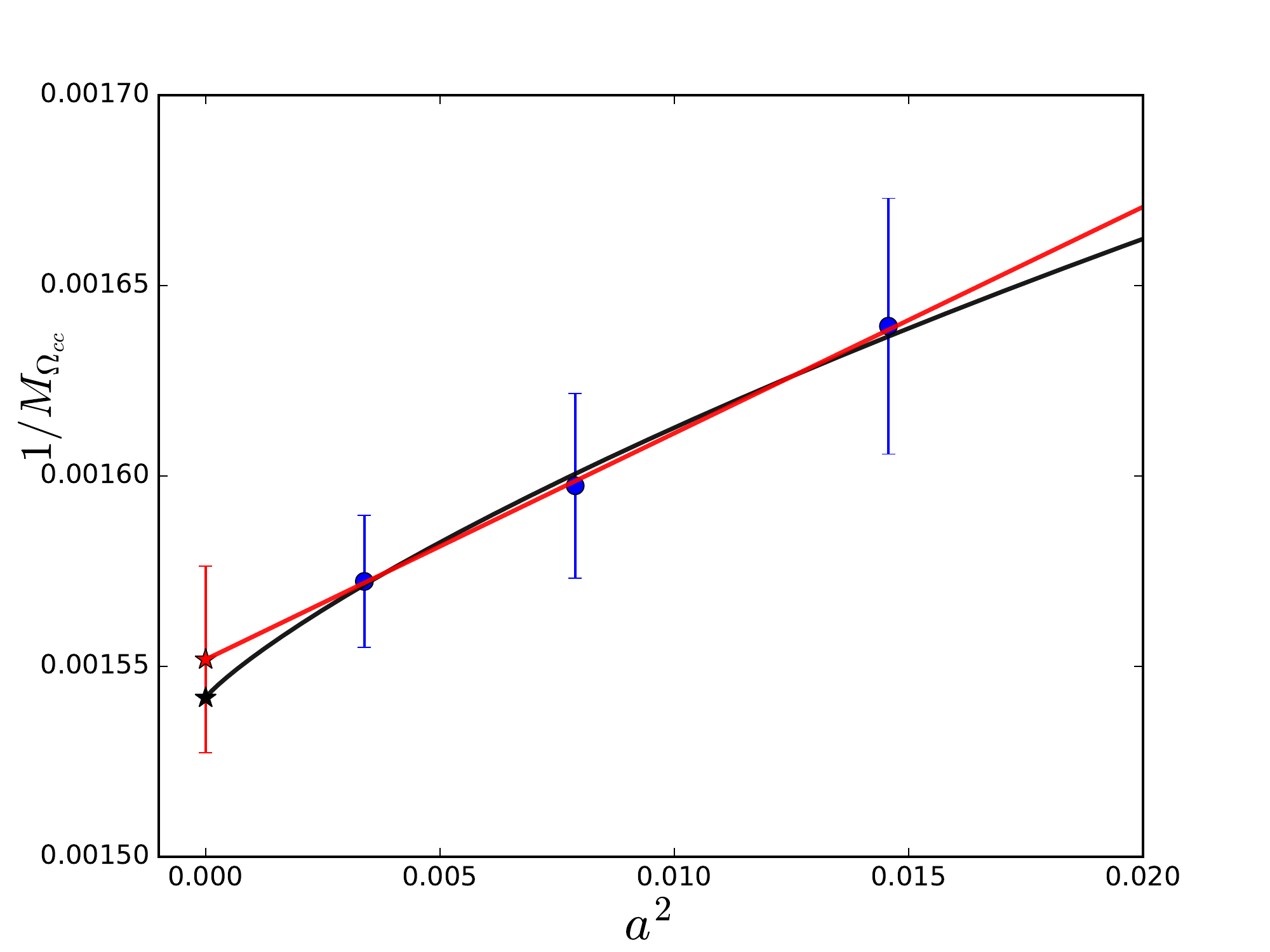}\\
\vspace*{-0.09in}
\caption{Fit of $1/M_{\Omega_{cc}(1/2^+)}$ with forms $A + B a^2$ and $A + C a^2\,log(a)$. Blue points with errorbar are data points and red and black lines are fitted lines with these two forms respectively. These fit results are showing that with this data set, within the $a^2$ range under consideration, it is not possible to quantitatively discern the $log$ term from the $a^2$ term.}
\eef{otherplot}


\vspace*{0.5in}

\subsection{Effective mass plots showing differences between the extracted masses obtained from HQET-based and relativistic operators at various quark masses}

Following plots present a comparison of effective masses extracted from the spin-1/2 projected HQET-based operator $[(q_1^T C\gamma_{i} q_1) q_2]$ and the relativistic operator $[(q_1^T C\gamma_{5} q_2) q_1]$ with $q_2$ set to strange quark, while $q_1$ refers to bottom (above), charm (middle) and light (below) quarks.
When $q_1 \equiv b$, they can be seen to be statistically equivalent, whereas with $q_1$ set to light and charm quark, statistically significant differences can be seen between the two estimates. This indicates there are strong relativistic effects from the charm quarks within the doubly charmed baryons and hence they may not be treated within an HQET-based picture.

\newpage

\bef[tbh]
\centering
\includegraphics*[scale=0.35]{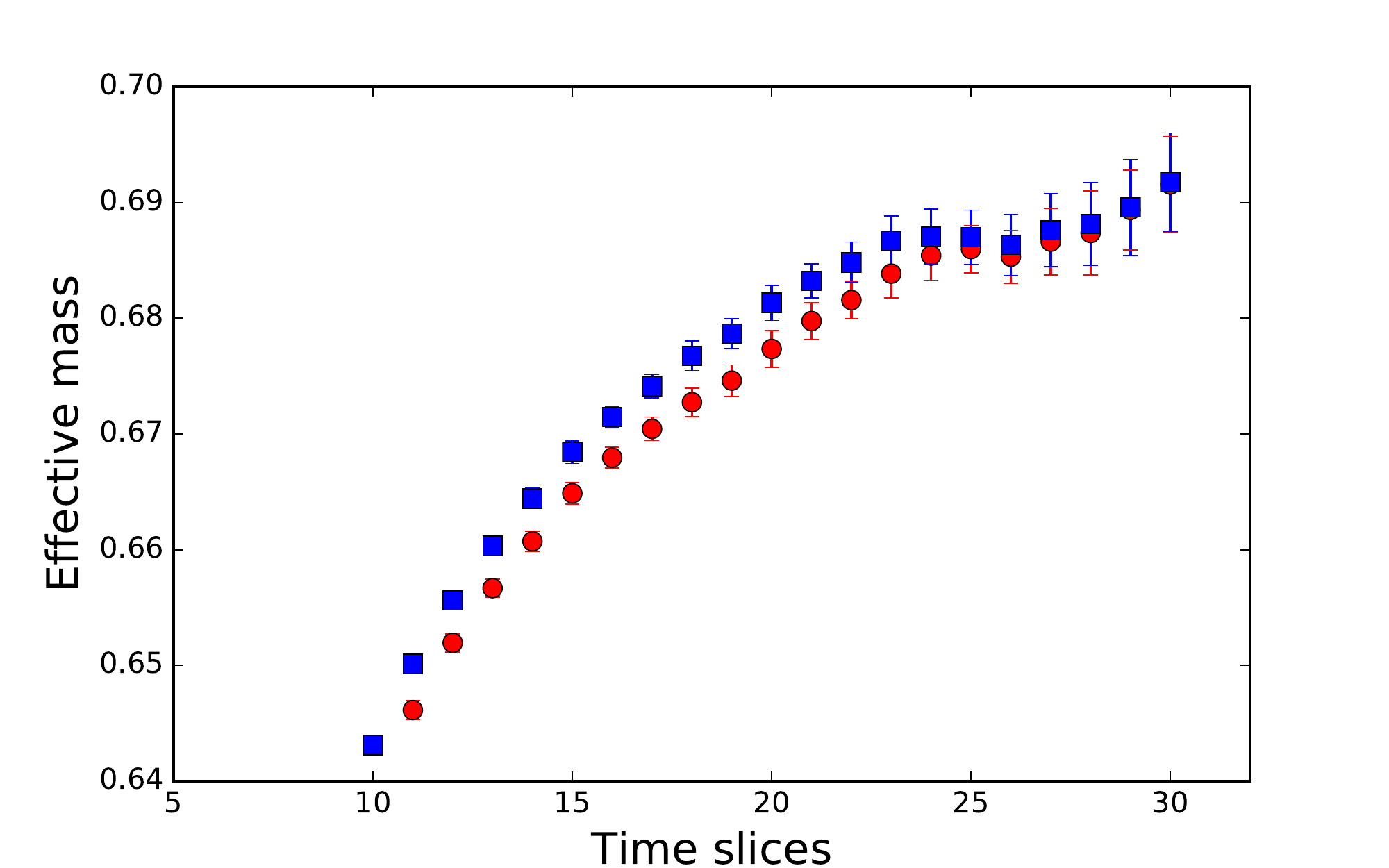}\\
\includegraphics*[scale=0.35]{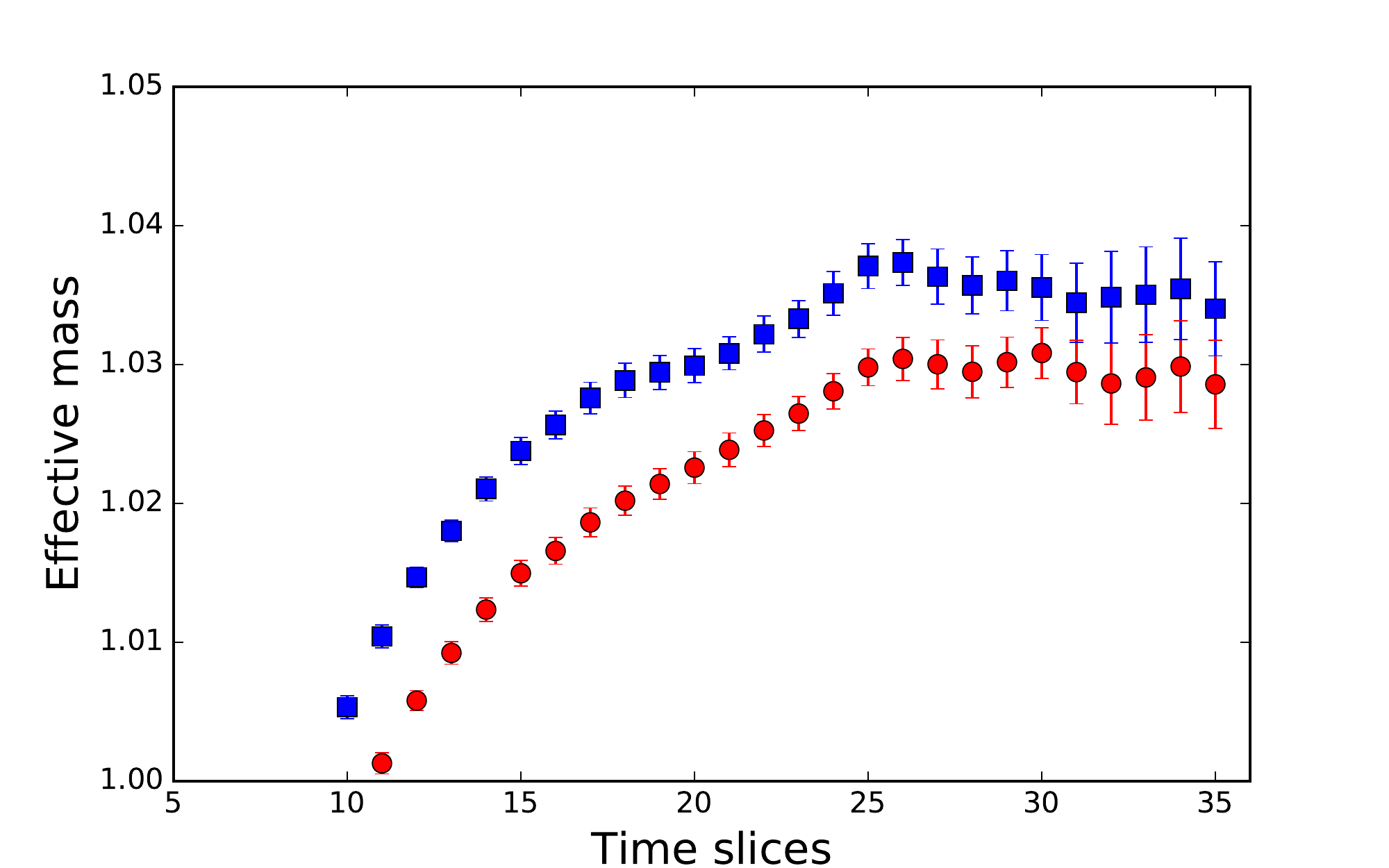}\\
\includegraphics*[scale=0.35]{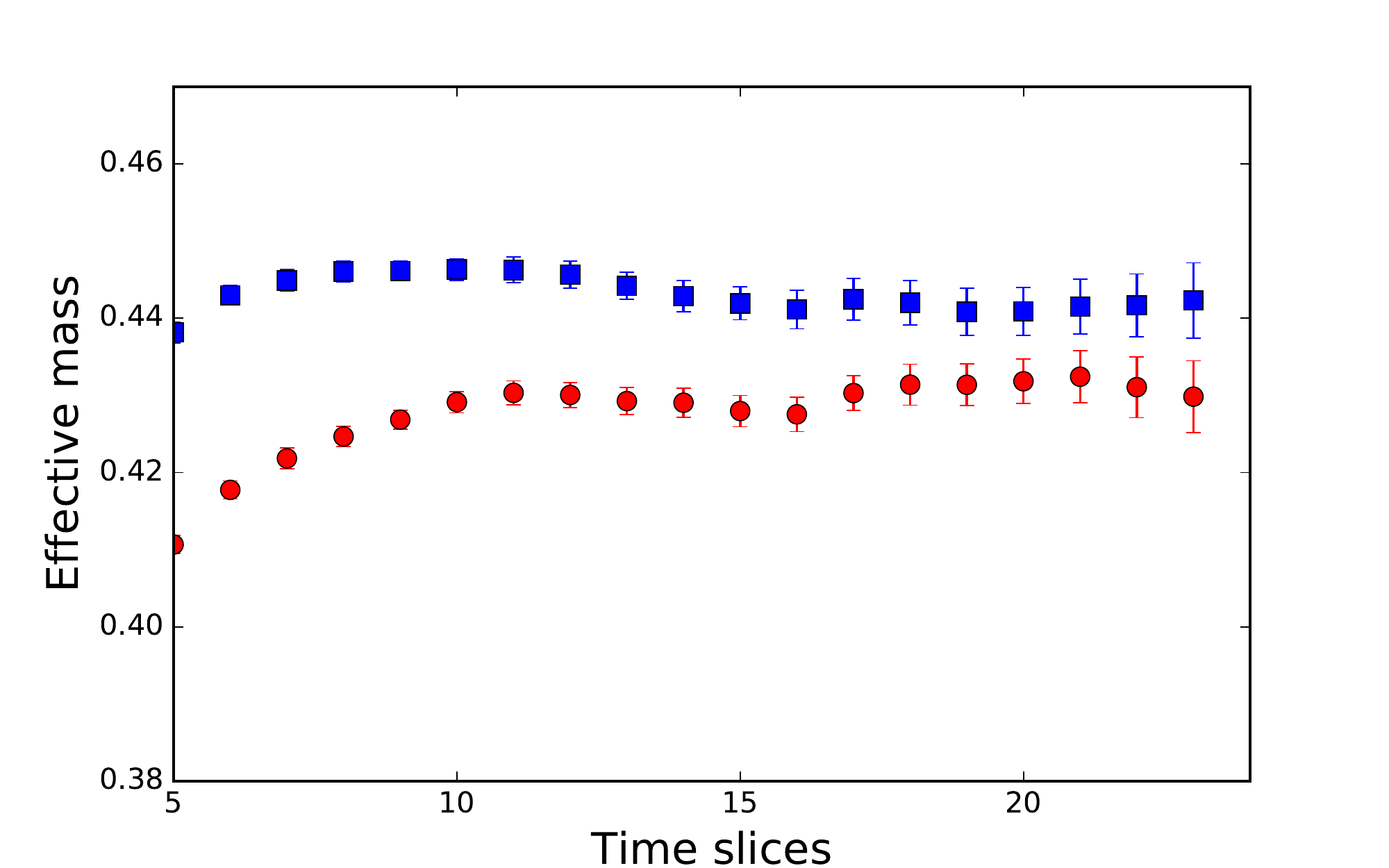}
\vspace*{-0.09in}
\caption{(Top) (Top) Effective mass plot of the ground state of $\Omega_{QQs}(1/2^{+})$ baryons corresponding to the relativistic (red circle) and HQET-based operators (blue square) on our fine lattice. The top figure is for the bottom quark ($bbs$), middle one is for the charm quark ($ccs$) and the bottom figure ($uus$) is for the quark mass $u$ corresponding to pion mass at $\sim$ 650 MeV. Strange quark mass is physical. In each figure, units in $y$-axis are arbitrary, corresponding to that quark mass. In physical units, differences of the central values of effective masses between these two operators are about 0, 25 and 35 MeV, respectively.}
\eef{hqet_test}

\end{document}